\documentclass[aps,prl,reprint,superscriptaddress]{revtex4-1}

\usepackage{graphicx}
\usepackage{float}
\usepackage{dcolumn}
\usepackage{bm}
\usepackage{amsmath}

\begin{document}


\title{Tuning Perpendicular Magnetic Anisotropy by Oxygen Octahedral Rotations in (La$_{1-x}$Sr$_{x}$MnO$_{3}$)/(SrIrO$_{3}$) Superlattices}

\author{Di Yi} \email{diyi@stanford.edu}
\affiliation{Geballe Laboratory for Advanced Materials, Stanford University, Stanford, California 94305, USA}
\affiliation{Department of Applied Physics, Stanford University, Stanford, California 94305, USA}

\author{Charles L. Flint}
\affiliation{Geballe Laboratory for Advanced Materials, Stanford University, Stanford, California 94305, USA}
\affiliation{Department of MSE, Stanford University, Stanford, California 94305, USA}

\author{Purnima P. Balakrishnan}
\affiliation{Geballe Laboratory for Advanced Materials, Stanford University, Stanford, California 94305, USA}
\affiliation{Department of Physics, Stanford University, Stanford, California 94305, USA}

\author{Krishnamurthy Mahalingam}
\affiliation{Materials and Manufacturing Directorate, Air Force Research Laboratory, Wright-Patterson AFB, Ohio 45433, USA}

\author{Brittany Urwin}
\affiliation{Materials and Manufacturing Directorate, Air Force Research Laboratory, Wright-Patterson AFB, Ohio 45433, USA}

\author{Arturas Vailionis}
\affiliation{Geballe Laboratory for Advanced Materials, Stanford University, Stanford, California 94305, USA}

\author{Alpha T. N'Diaye}
\affiliation{Advanced Light Source, Lawrence Berkeley National Laboratory, Berkeley, California 94720, USA}

\author{Padraic Shafer}
\affiliation{Advanced Light Source, Lawrence Berkeley National Laboratory, Berkeley, California 94720, USA}

\author{Elke Arenholz}
\affiliation{Advanced Light Source, Lawrence Berkeley National Laboratory, Berkeley, California 94720, USA}

\author{Yongseong Choi}
\affiliation{Advanced Photon Source, Argonne National Laboratory, Argonne, Illinois 60439, USA}

\author{Kevin H. Stone}
\affiliation{SSRL, SLAC National Accelerator Laboratory, Menlo Park, California 94025, USA}

\author{Jiun-Haw Chu}
\affiliation{Geballe Laboratory for Advanced Materials, Stanford University, Stanford, California 94305, USA}
\affiliation{Department of Applied Physics, Stanford University, Stanford, California 94305, USA}
\affiliation{SIMES, SLAC National Accelerator Laboratory, Menlo Park, California 94025, USA}

\author{Brandon M. Howe}
\affiliation{Materials and Manufacturing Directorate, Air Force Research Laboratory, Wright-Patterson AFB, Ohio 45433, USA}

\author{Jian Liu}
\affiliation{Department of Physics and Astronomy, University of Tennessee, Knoxville, Tennessee 37996, USA}

\author{Ian R. Fisher}
\affiliation{Geballe Laboratory for Advanced Materials, Stanford University, Stanford, California 94305, USA}
\affiliation{Department of Applied Physics, Stanford University, Stanford, California 94305, USA}
\affiliation{SIMES, SLAC National Accelerator Laboratory, Menlo Park, California 94025, USA}

\author{Yuri Suzuki}
\affiliation{Geballe Laboratory for Advanced Materials, Stanford University, Stanford, California 94305, USA}
\affiliation{Department of Applied Physics, Stanford University, Stanford, California 94305, USA}

\date{\today}

\begin{abstract}
Perpendicular magnetic anisotropy (PMA) plays a critical role in the development of spintronics, thereby demanding new strategies to control PMA. Here we demonstrate a conceptually new type of interface induced PMA that is controlled by oxygen octahedral rotation. In superlattices comprised of La$_{1-x}$Sr$_{x}$MnO$_{3}$ and SrIrO$_{3}$, we find that all superlattices (0$\leq$\emph{x}$\leq$1) exhibit ferromagnetism despite the fact that La$_{1-x}$Sr$_{x}$MnO$_{3}$ is antiferromagnetic for \emph{x}$>$0.5. PMA as high as 4$\times$10$^6$ erg/cm$^3$ is observed by increasing \emph{x} and attributed to a decrease of oxygen octahedral rotation at interfaces. We also demonstrate that oxygen octahedral deformation cannot explain the trend in PMA. These results reveal a new degree of freedom to control PMA, enabling discovery of emergent magnetic textures and topological phenomena.
\end{abstract}

\maketitle

The manipulation of spin in magnetic systems has been of great interest as it has given rise to a rich spectrum of magnetic ground states and has enabled high performance magnetic memory and logic devices. Magnetic anisotropy (MA), which describes the tendency of magnetic moment vectors to prefer specific directions, plays an important role in determining these ground states. In particular, perpendicular magnetic anisotropy (PMA), where the magnetic moments preferentially point perpendicular to the film plane, has been of fundamental and technological interest. PMA is critical to realizing devices with high density, high stability and low energy consumption \cite{Man,Ike,Liu,Emo}. Competition between PMA and other mechanisms, such as the Dzyaloshinskii-Moriya interaction, leads to rich spin textures such as chiral domain walls or magnetic skyrmions and novel topological phenomena \cite{Nag}. Therefore it is important to explore strategies to control PMA in various ferromagnetic materials.

Transition metal oxides (TMOs) of the form ABO$_{3}$ are interesting candidates. In this family of materials, magnetic interactions are dictated largely by bonds among transition metal cations and oxygen anions (B-O bonds). The magnetocrystalline anisotropy is determined by the spin-orbit coupling (SOC) and anisotropy of structure that includes B-O bond distances and angles. Modifications in B-O bond distances and angles are described in terms of oxygen octahedral deformation (OOD) and rotation (OOR) \cite{Ron,Gla1}. Previously PMA of TMO films has been shown to be tunable by epitaxial strain that leads to OOD in the form of unequal B-O bond distances between in-plane and out-of-plane directions \cite{Wu}. Recent studies have revealed the important role of OOR in tuning MA in terms of bond angles \cite{Lia,Bos,Kan}. However, in both approaches, the tuning of PMA is attributed to the modification of crystal structure within the ferromagnetic TMOs. Recently magnetism in TMOs with strong SOC has been shown to be largely determined by anisotropic exchange coupling \cite{Jac,Kim}. Therefore PMA may be induced at interfaces between ferromagnetic TMOs and TMOs with strong SOC via interfacial and in-plane exchange interactions, thereby suggesting a conceptually different strategy to tuning PMA.

In this letter, we demonstrate such interface induced PMA where the strength of the PMA is determined by the degree of OOR at the 3\emph{d}-5\emph{d} TMOs interfaces. Here we study superlattices (SLs) comprised of 3\emph{d} TMOs La$_{1-x}$Sr$_{x}$MnO$_{3}$ (LSMO, 0$\leq$\emph{x}$\leq$1) and 5\emph{d} TMOs SrIrO$_{3}$ (SIO). We find that all SLs exhibit ferromagnetism despite the fact that LSMO is antiferromagnetic for \emph{x}$>$0.5 \cite{Tok}. The ferromagnetic ground state evolves as a function of \emph{x} with PMA increasing from around zero to as high as 4$\times$10$^6$ erg/cm$^3$ with increasing \emph{x}, contrary to the trend expected from OOD. We show that the doping (\emph{x}) dependent OOR is closely correlated with PMA, implying the critical role of interfacial bond angles. This strategy can be extended to a wider class of heterostructures comprised of other 3\emph{d} and 4\emph{d}/5\emph{d} TMOs.

Epitaxial (001) [(LSMO)$_{1}$(SIO)$_{1}$]$_{25}$ SLs, consisting of 25 repeats of 1 unit cell of LSMO and 1 unit cell of SIO, have been grown on cubic SrTiO$_{3}$ substrates to explicitly probe the interfacial effect. In this study, \emph{x} is chosen to be 0, 0.3, 0.5, 0.8 and 1. We have also grown single layer films of LSMO and SIO, SLs with the same ratio of LSMO to SIO but larger periodicity and [(La$_{0.3}$Ca$_{0.7}$MnO$_{3}$)$_{1}$(SIO)$_{1}$]$_{25}$ SLs on SrTiO$_{3}$ as references.  It is noted that the lattice constant of LSMO (pseudo-cubic a$_{pc}$ increases from $\sim$3.80{{\AA}} to $\sim$3.905{{\AA}} with decreasing \emph{x}) \cite{Mit,Chm} is smaller than that of SrTiO$_{3}$ (a=3.905{{\AA}}) while SIO (a$_{pc}$=3.975{{\AA}}) has a larger lattice constant \cite{Lon}. Structural characterization is shown in the Supplemental Material (Fig. S1). The high degree of crystallinity, epitaxy, designed periodicity and interface abruptness of the SLs are evident in the scanning transmission electron microscopy images and the x-ray diffraction (XRD) patterns. Reciprocal space mapping confirms that the SLs are coherently strained by the substrates. The mean out-of-plane lattice constants (\emph{c}) of the SLs, determined from the (002) Bragg peak of XRD patterns, are larger than that of SrTiO$_{3}$ and linearly decrease as \emph{x} increases.

\begin{figure}\vspace{-0pt}
\centering
\includegraphics[width=8.6cm]{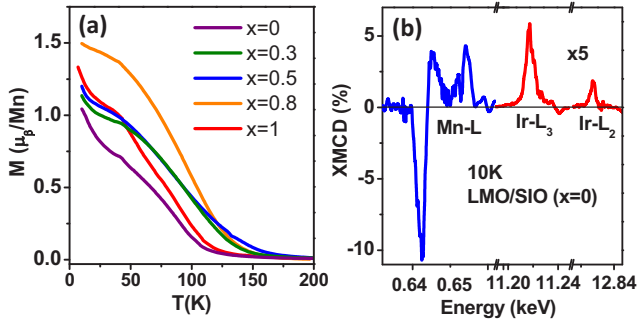}
\caption{\label{Magnetism} \textbf{Ferromagnetism in SLs}. (a) Temperature dependence of magnetization for SLs. (b) XMCD spectra of SLs (\emph{x}=0) at both the Mn and Ir \emph{L$_{2,3}$} edges measured under the same experimental conditions.}
\end{figure}

The first interesting observation is that all SLs exhibit ferromagnetic behavior, which is not expected since SIO is paramagnetic \cite{Lon} and LSMO is antiferromagnetic when \emph{x}$>$0.5 (Supplemental Material, Fig. S2) \cite{Tok,Kon,Mats,Rit,Cho,Gib}. Fig. \ref{Magnetism}(a) shows the temperature dependence of magnetization for the series of SLs. These SLs were field cooled in 7T and then measured during the warming process with 0.1T applied perpendicular to the film plane, which is significantly smaller than the saturation field. The varying magnitude of magnetization for different \emph{x} in Fig. \ref{Magnetism}(a) is a reflection of PMA as will be discussed below. To identify the origin of moments, we performed element-selective x-ray magnetic circular dichroism (XMCD) measurements at the \emph{L$_{2,3}$} edges of both Mn and Ir. Representative results are shown in Fig. \ref{Magnetism}(b) and are similar to spectra from previous reports \cite{Yi,Nic}. These results reveal several aspects of magnetism in the SLs. First of all, magnetization of LSMO/SIO SLs is mainly due to LSMO spin moments by comparing the magnitude of the XMCD signals to references (Supplemental Material, Fig. S3). Secondly, the small induced magnetization in SIO couples antiparallel to the moments of LSMO \cite{Yi,Nic}. Thirdly, the same sign and comparable amplitude of the XMCD signal at the Ir \emph{L$_{2}$} and \emph{L$_{3}$} edges indicate a large contribution from orbital moments \cite{Yi,Nic}. Finally, the emergent ferromagnetism in Sr-rich SLs is due to a charge transfer effect at the interface \cite{Nic}, which has been observed in other systems \cite{Yam,Gar1,Gar2}. The transfer of electrons to Mn from Ir orbitals is evident in the x-ray absorption spectroscopy (XAS) data. XAS of the Mn \emph{L$_{2,3}$} edges shows a significant increase of the \emph{L$_{3}$/L$_{2}$} ratio in \emph{x}=1 SLs compared to SrMnO$_{3}$ films (Supplemental Material, Fig. S3), thus indicating a transfer of electrons from Ir$^{4+}$ to Mn$^{4+}$ \cite{Var,kob}.

\begin{figure*}[t]\vspace{-0pt}
\centering
\includegraphics[width=17.2cm]{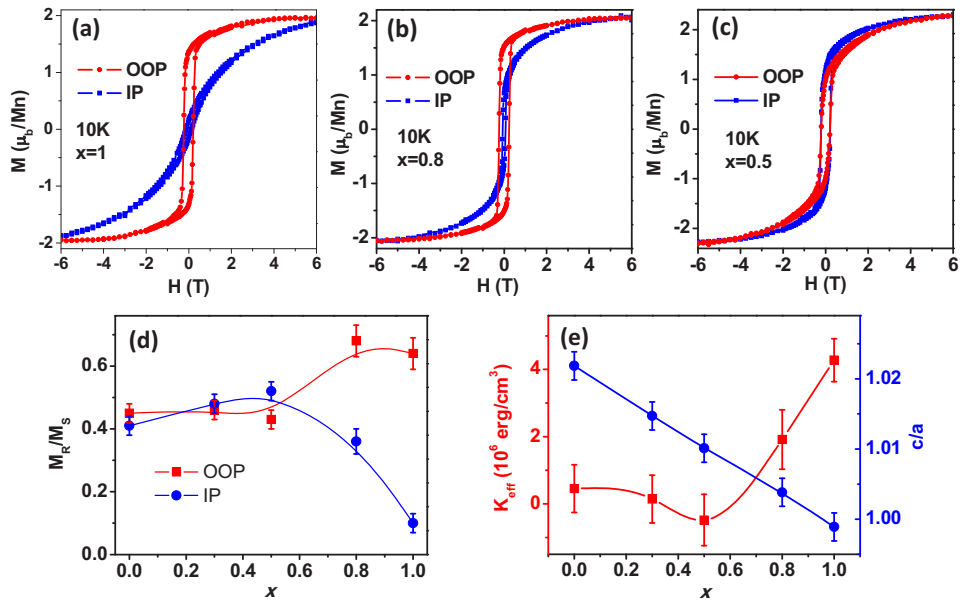}
\caption{\label{PMA} \textbf{Perpendicular magnetic anisotropy in SLs.} (a)-(c) Magnetic hysteresis loops along in-plane (IP) and out-of-plane (OOP) directions of SLs with (a) \emph{x}=1, (b) \emph{x}=0.8 and (c) \emph{x}=0.5. (d) The \emph{x} dependence of normalized remnant magnetizations along IP and OOP directions. (e) Comparison of \emph{x} dependence of PMA constatnt (\emph{K$_{eff}$}) and mean tetragonal distortion (\emph{c/a}).}
\end{figure*}

In order to show the evolution of PMA, magnetization loops were measured along both in-plane ([100]) and out-of-plane ([001]) directions and the representative results are shown in Fig. \ref{PMA}(a)-(c) for SLs with \emph{x}=1, 0.8 and 0.5. It is noted that magnetization of LSMO films on SrTiO$_{3}$ align in the film plane due to tensile strain \cite{Suz}. However the large difference in the in-plane and out-of-plane hysteresis loops of the SLs with \emph{x}=1 indicates a strong PMA \cite{Nic}. This emergent PMA exhibits different behaviors depending on \emph{x}: the difference in the in-plane and out-of-plane loops decreases with decreasing \emph{x} for 0.5$\leq$\emph{x}$\leq$1 and effectively disappears for \emph{x}$\leq$0.5 (Supplemental Material, Fig. S4). We have also measured magnetization of [(LSMO)$_{3}$(SIO)$_{3}$]$_{10}$ (\emph{x}=0.3) SLs versus magnetic field. The differences in magnetic behavior of [1,1] and [3,3] SLs confirm the PMA to be an interface effect (Supplemental Material, Fig. S4).

Fig. \ref{PMA}(d) shows the \emph{x} dependence of the remnant (M$_{R}$) to saturation (M$_{S}$) magnetization ratio along the two directions, revealing that the magnetic easy axis shifts away from the out-of-plane direction as \emph{x} decreases. To quantify the strength of PMA, we estimate an effective anisotropy constant (\emph{K$_{eff}$}) associated with PMA by the area enclosed between the out-of-plane and in-plane magnetization curves (Supplemental Material, Fig. S4) \cite{Joh}. When the magnetization is preferably oriented perpendicular to the plane, \emph{K$_{eff}$} is positive by this definition and its magnitude reflects the strength of PMA. The largest \emph{K$_{eff}$} is found to be as high as 4$\times$10$^6$ erg/cm$^3$ in \emph{x}=1 SLs (Fig. \ref{PMA}(e)), which is one order of magnitude higher than that of the strained La$_{0.7}$Sr$_{0.3}$MnO$_{3}$ \cite{Wu} and comparable to SrRuO$_{3}$ \cite{Kos}. This dramatic enhancement of magnetic anisotropy energy reveals the important role of the SIO layer and its interaction with the manganite layer. As \emph{x} decreases, \emph{K$_{eff}$} decreases significantly and disappears for \emph{x}$\leq$0.5. It can be concluded from Fig. \ref{PMA} that a strong PMA is induced by interfacing LSMO with SIO for large values of \emph{x}. Moreover, this interfacial PMA can be widely tuned as a function of A-site substitution (\emph{x}). Since exchange interactions are sensitive to the bond distances and angles that are determined by the oxygen network, a comprehensive probe into the local structural distortions is essential in revealing the key factors that control PMA at the interfaces.

\begin{figure}[t]
\centering
\includegraphics[width=8.6cm]{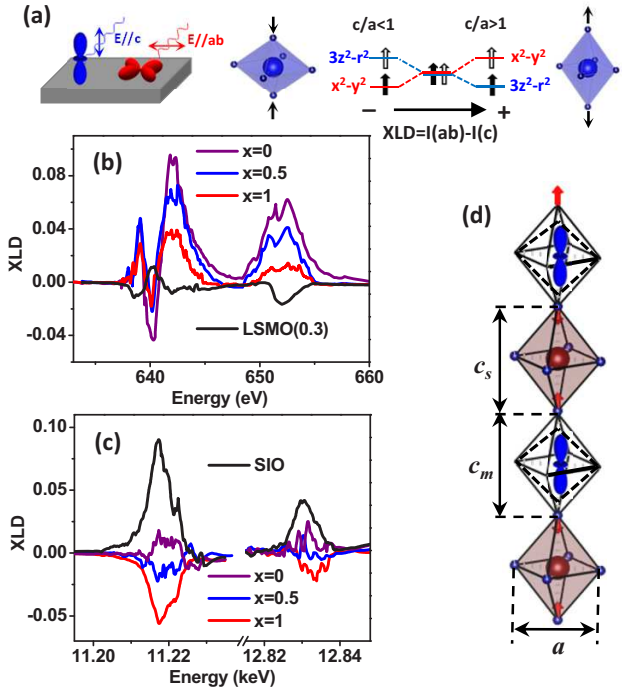}
\caption{\label{XLD} \textbf{Oxygen octahedral deformation in SLs.} (a) Sketch of the correlation between XLD and OOD. (b), (c) XLD spectra of SLs and reference (La$_{0.7}$Sr$_{0.3}$MnO$_{3}$ and SrIrO$_{3}$) films at the Mn and Ir \emph{L$_{2,3}$} edges at room temperature. (d) Sketch of OOD in SLs.}
\end{figure}

We first study the role of OOD, which depends on the tetragonal distortion that can be described by the \emph{c/a} ratio. The OOD of MnO$_{6}$ (IrO$_{6}$) in our SLs is described by the \emph{c$_{m}$/a} (\emph{c$_{s}$/a}) ratio as shown in Fig. \ref{XLD}(d). To probe them individually, we measured x-ray linear dichroism (XLD) at the Mn and Ir \emph{L$_{2,3}$} edges at room temperature \cite{Aru}. As shown in Fig. \ref{XLD}(a), the different linearly polarized x-rays excite electrons into different \emph{d} orbitals. Here the XLD is calculated as the intensity difference (I(ab)-I(c)) between the normalized XAS spectra measured with in-plane (E//ab) and out-of-plane (E//c) polarizations. The octahedral distortion with \emph{c/a}$<$1 (\emph{c/a}$>$1) leads to more empty out-of-plane (in-plane) \emph{d} states and thus negative (positive) XLD \cite{Teb,Pes,Pesq}. Fig. \ref{XLD}(b) and (c) show XLD of SLs (\emph{x}=0, 0.5, 1) and reference films (La$_{0.7}$Sr$_{0.3}$MnO$_{3}$ and SIO on SrTiO$_{3}$). For SLs with \emph{x}=1, the XLD spectra clearly show opposite signs compared to the reference films at both the Mn and Ir edges. The results reveal an interface-driven OOD that is probably due to the shift of the apical oxygen anions (Fig. \ref{XLD}(d)) \cite{Lep,San}. As \emph{x} decreases, Fig. \ref{XLD}(b) shows that the XLD at the Mn edges increases and Fig. \ref{XLD}(c) shows that the XLD at the Ir edges changes from negative to positive. These results reveal an increase of both \emph{c$_{m}$/a} and \emph{c$_{s}$/a} as \emph{x} decreases in our SLs (Supplemental Material, Fig. S5), consistent with the trend of the mean \emph{c/a} ratio obtained from XRD. Previous studies have demonstrated that PMA in manganites is enhanced as \emph{c$_{m}$/a} increases \cite{Wu,Kimu}, which lifts the e$_{g}$ orbital degeneracy that is coupled to spin by SOC \cite{Hag}. Moreover, an increase of \emph{c$_{s}$/a} in Sr$_{n+1}$Ir$_{n}$O$_{3n+1}$ changes the superposition of the t$_{2g}$ orbitals in the J$_{eff}$=1/2 state and thus the anisotropic exchange interactions, leading to a spin-flop transition from in-plane to out-of-plane \cite{Jac,Kim,Liu2}. Therefore one would expect the enhancement of PMA in our SLs as \emph{c/a} increases if OOD is the dominant mechanism. Fig. \ref{PMA}(e) compares the \emph{x} dependence of \emph{K$_{eff}$} with that of the mean \emph{c/a}. As \emph{c/a} linearly increases with decreasing \emph{x}, the PMA constant actually decreases and remains mostly unchanged for \emph{x}$\leq$0.5. Therefore OOD cannot explain the evolution of PMA here.

\begin{figure}[b]
\centering
\includegraphics[width=8.6cm]{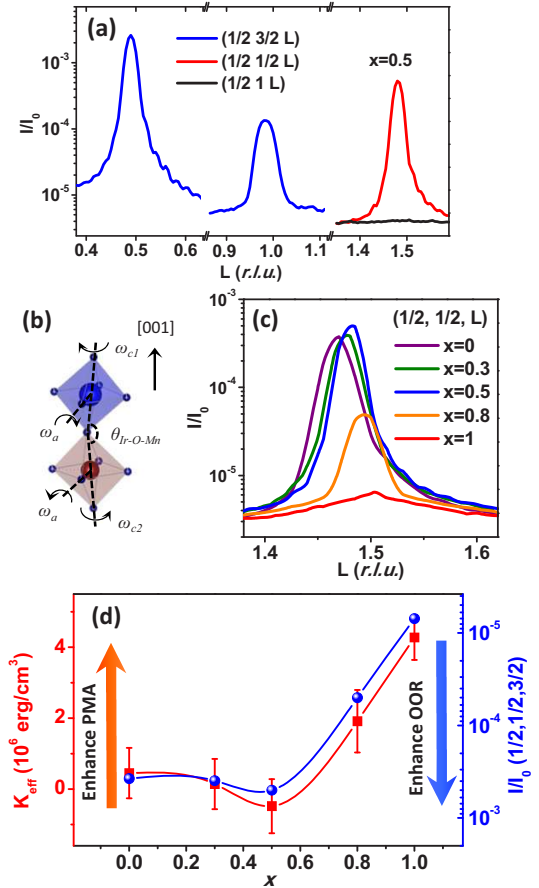}
\caption{\label{XRD} \textbf{Oxygen octahedral rotation in SLs.} (a) L scan along (00L) crystal truncation rod for a SL (\emph{x}=0.5). (b) Sketch of OOR geometry in SLs. (c) Normalized (1/2, 1/2, 3/2) peaks for SLs with different \emph{x}. Intensity is normalized by the corresponding (002) Bragg peak. (d) Comparison of the \emph{x} dependence of PMA (\emph{K$_{eff}$}) and OOR. }
\end{figure}

Another possible mechanism to account for the trend of PMA is OOR. OOR can be described by rotations along three orthogonal directions \cite{Ron}. It has been demonstrated that OOR gives rise to unique XRD intensity profiles of half-order Bragg peaks (\emph{h}/2, \emph{k}/2, \emph{l}/2) \cite{Gla2,May1,May2}. Half-order peaks with certain combinations of odd/even \emph{h}, \emph{k} and \emph{l} reveal rotation phases along particular directions. Moreover, the intensity of the half-order peak is proportional to the corresponding rotation angle. As shown in Fig. \ref{XRD}(b), OOR of our SLs can be viewed as rotations along two symmetric in-plane axes \emph{\textbf{a}} (\emph{$\omega$$_{a}$}) and the out-of-plane axis \emph{\textbf{c}} ($\omega$$_{c1}$, $\omega$$_{c2}$). Here we measured a group of half-order peaks (1/2, 1/2, \emph{l}/2), (1/2, 1, \emph{l}/2) and (1/2, 3/2, \emph{l}/2). Representative results are shown in Fig. \ref{XRD}(a) (\emph{x}=0.5). Based on previous studies \cite{Gla2,May1,May2,Zha,Gru}, we can identify the dominant rotation contributions for each peak. More specifically, the observation of the (1/2, 1/2, 3/2) peak corresponds to out-of-phase rotations along \emph{\textbf{a}}, consistent with the absence of the (1/2, 1, 3/2) peak that corresponds to in-phase rotations along \emph{\textbf{a}}. The large intensity of the (1/2, 3/2, 1/2) peak indicates an out-of-phase rotation along \emph{\textbf{c}} \cite{May1}. The (1/2, 3/2, 1) peak is also observed and the intensity is about one order of magnitude lower than that of (1/2, 3/2, 1/2). This peak is due to the coherent modulation of OOR along \emph{\textbf{c}} in LSMO ($\omega$$_{c1}$) and SIO ($\omega$$_{c2}$) in one SL period. Observation of thickness fringes in Fig. \ref{XRD}(a) implies that OOR is correlated across interfaces. It is noted that similar long-range modulation of half-order peaks has been observed in other SLs \cite{May2,Zha}. Further analyses of OOR are included in the Supplemental Material (section 6).

To probe the correlation between OOR and PMA, we track the \emph{x} dependence of the half-order peak intensities. Fig. \ref{XRD}(c) shows the normalized intensities of (1/2, 1/2, 3/2), corresponding to rotations along \emph{\textbf{a}} ($\omega$$_{a}$), for SLs with different values of \emph{x}. This peak is almost negligible for SLs with \emph{x}=1, indicating a suppression of such rotation. As \emph{x} decreases, the peak intensity increases by two orders of magnitudes and saturates for \emph{x}$\leq$0.5. Although further measurements are required to quantify the rotation angles, the \emph{x} dependence of peak intensity closely correlates with that of PMA as shown in Fig. \ref{XRD}(d). Both the PMA constant (\emph{K$_{eff}$}) and the (1/2, 1/2, 3/2) peak intensity exhibit a similar doping dependence for 0.5$\leq$\emph{x}$\leq$1 and saturation for 0$\leq$\emph{x}$\leq$0.5. We also track the (1/2, 3/2, 1/2) peak that corresponds to rotations along \textbf{\emph{c}} (Supplemental Material, Fig. S6). The intensity increases as \emph{x} decreases but does not show a clear saturation behavior. Therefore it is likely that PMA is largely correlated to rotations along \emph{\textbf{a}}. Additional evidence for this correlation is found in [(La$_{0.3}$Ca$_{0.7}$MnO$_{3}$)$_{1}$(SIO)$_{1}$]$_{25}$ SLs where rotations along \emph{\textbf{c}} cannot be correlated with PMA (see Supplemental Material, section 7). Since the interfacial Ir-O-Mn bond angles are determined by rotations along \textbf{\emph{a}} ($\omega$$_{a}$), the results imply that straight Ir-O-Mn bonds ($\sim$180$^\circ$) stabilize an out-of-plane magnetic easy axis with a strong PMA. As rotations along \textbf{\emph{a}} ($\omega$$_{a}$) increase and Ir-O-Mn bond angles decrease, the easy axis tilts away from out-of-plane and the PMA decreases.

Results in Fig. \ref{XRD} reveal that the interfacial bond angle is a key parameter to tuning PMA. Here variations of Ir-O-Mn bond angles affect the orbital overlap of the two cations bridged via oxygen, thus effectively changing the exchange interactions across interfaces along the out-of-plane direction. Although magnetization of the SLs is dominated by LSMO, the PMA can be affected by both SIO and LSMO sublayers. The spin-flop from in-plane to out-of-plane has been observed in layered-iridates Sr$_{n+1}$Ir$_{n}$O$_{3n+1}$ due to the variation of anisotropic exchange interactions \cite{Kim} and may have similar origins to the tuning of PMA in our SLs. Moreover, straight bonds have been shown to enhance the electron hopping between neighboring Mn cations in LSMO films and thus gives rise to a uniaxial magnetic easy axis \cite{Lia}, which may also contribute to our observations. In any case, our results distinctly demonstrate that the interface induced PMA is strongly correlated with the degree of OOR. This conclusion obtained from LSMO/SIO interfaces suggests that heterostructures based on ferromagnetic 3\emph{d} and 4\emph{d}/5\emph{d} TMOs are promising for generating a tunable PMA by carefully designing OOR at interfaces.

In conclusion, we have demonstrated a significant PMA induced at 3\emph{d}-5\emph{d} TMOs interfaces. We have found that the strength of PMA is associated with interfacial bond angles that are controlled by OOR, which dominates over OOD. Our results underscore the importance of not only the shape but also the connectivity of oxygen octahedra in determining the emergent interfacial phenomena. This new strategy to control PMA can potentially be extended to a wider class of materials. The tunable interfacial PMA, combined with the broken inversion symmetry and SOC effect, also reveals the ideal candidates to search for emergent magnetic textures and topological phenomena in TMOs \cite{Mat}.

\begin{acknowledgments}
The work at Stanford University performed by D.Y. and Y.S is supported by the Air Force Office of Scientific Research (AFOSR) under Grant No. FA9550-16-1-0235 and was initially supported by the Vannevar Bush Faculty Fellowship program, sponsored by the Basic Research Office of the Assistant Secretary of Defense for Research and Engineering under the Office of Naval Research Grant No. N00014-15-1-0045. C.L.F. and P.P.B. assisted in spectroscopy measurements. C.L.F. is funded by the Director, Office of Science, Office of Basic Energy Sciences of the U.S. Department of Energy under contract No. DE-SC0008505. P.P.B. is funded by the the National Science Foundation under DMR-1402685. J.-H.C. and I.R.F., and synthesis of the SrIrO$_{3}$ target, are supported by the DOE, Office of Basic Energy Sciences under contract DE-AC02-76SF00515. J.L. acknowledges support from the Science Alliance Joint Directed Research and Development Program and the Transdisciplinary Academy Program at the University of Tennessee. J.L. also acknowledges support by the DOD-DARPA under Grant No. HR0011-16-1-0005. This research used resources of the Advanced Light Source, which is a DOE Office of Science User Facility under contract no. DE-AC02-05CH11231. Use of the Stanford Synchrotron Radiation Lightsource, SLAC National Accelerator Laboratory, is supported by the U.S. Department of Energy, Office of Science, Office of Basic Energy Sciences under Contract No.DE-AC02-76SF00515. Use of the Advanced Photon Source, an Office of Science User Facility operated for the U.S. DOE, OS by Argonne National Laboratory, was supported by the U.S. DOE under Contract No. DE-AC02-06CH11357.
\end{acknowledgments}

\bibliography{PMASIOLSMO}

\end{document}